\documentclass{ws-mpla}
\usepackage[super]{cite}
\usepackage{graphicx}
\begin{document}

\markboth{R.Delbourgo \& P.D.Stack}
{Relativity for $N$ Properties}

\catchline{}{}{}{}{}

\title{General Relativity for $N$ Properties.}

\author{\footnotesize ROBERT DELBOURGO and PAUL D STACK}

\address{School of Physical Sciences, University of Tasmania, GPO Box 37\\
Hobart, TAS 7005, Australia
\\
bob.delbourgo@utas.edu.au, pdstack@utas.edu.au}

\maketitle

\pub{Received (Day Month Year)}{Revised (Day Month Year)}

\begin{abstract}
We determine the coefficients of the terms multiplying the gauge fields, gravitational field and
cosmological term in a scheme whereby properties are characterized by $N$ anticommuting
scalar Grassmann variables. We do this for general $N$, using analytical methods; this obviates the 
need for our algebraic computing package which can become quite unwieldy as $N$ is increased.

\end{abstract}

\ccode{PACS Nos.: include PACS Nos.}

\section{Structure of the Lagrangian terms}	

Over the last few years we have developed a scheme whereby the gravity is unified
with other force fields by embracing them all in a supermetric which features spacetime
augmented by Lorentz scalar anticommuting coordinates $\zeta^\mu$. These $\zeta$ specify 
the characteristics or properties of an event\cite{paper1,paper2,paper3,paper4}, 
in addition to location and time $x^m$. This scheme\footnote{One can always make any gauge
field Lagrangian consistent with general relativity by ensuring invariance under
spacetime coordinate changes through the gravitational metric or vierbein but consistency 
does not mean proper unification.} resulted in a Lagrangian which
contained the Yang-Mills, gravitational  and cosmological terms, consistent with 
general coordinate invariance, but which pointed to the need for coupling constant
unification at an appropriately high length scale $l$ at least when chirality was involved. 
We were able to make progress for
the case of a small number of properties by using of an algebraic computation
package like Mathematica; this provided ample confirmation for the gauge invariance 
of the result by explicit computation of the super-Ricci scalar. The final answer was
dependent on a number of property curvature coefficients and the calculation became
progressively more difficult (and expensive in  machine time) as the number $N$ of 
properties got larger, in a factorial sense. In this letter we shall describe a way
of extracting the result for any $N$ by an analytical method which significantly 
obviates the need for extensive computer calculation and is a major advance in
tackling the practical case of $N=5$ (or $N$=10 if we distinguish chiralities)
which comprises the standard model.

The trick which leads to such an advance is based upon the Palatini form
of the Ricci superscalar $\cal R$:
\begin{equation}
{\cal R} = (-1)^{[L]}G^{MK}[(-1)^{[L][M]}{\Gamma_{KL}}^N{\Gamma_{NM}}^L
                - {\Gamma_{KM}}^N{\Gamma_{NL}}^L],
\end{equation}
where $M=m$ for spacetime or $M=\mu$ for graded property; the Christoffel symbol 
$\Gamma$ is defined by
\begin{equation}
2{\Gamma_{MN}}^K \equiv [(-1)^{[M][N]}G_{MR,N} + G_{NR,M} -
                              (-1)^{[R]([M]+[N])} G_{MN,R}](-1)^{[R]}G^{RK},
\end{equation}
all derivatives are left-sided and the square brackets denote the grading 
($0$ or $1$) of the appropriate index. The significant remark is that $\cal R$
formally possesses the structure 
$G^{KL}G^{MN}G^{RS}(\partial_.G_{..}\partial_.G_{..})_{KLMNRS}$ with the 
index subscripts $KLMNRS$ distributed between the derivatives and metrics
within the brackets ( ).
It then becomes an exercise in picking out typical terms which can contribute to the 
gravitational curvature, the gauge fields and the cosmological constant.

\section{The supermetric} 
As explained in previous work the metric (derived from triangular frame vectors)
which contains gravity and the gauge fields  (denoted by $A$ and ignoring coupling constants) 
has the following components:
\begin{eqnarray}
x-x~{\rm sector},~G_{mn}&=&g_{mn}C +l^2\bar{\zeta}(A_mA_n+A_nA_m)\zeta C'/2,\\
x-\zeta~{\rm sector},~G_{m\nu}&=&-il^2(\bar{\zeta}A_m)^{\bar{\nu}} C'/2,\\
\zeta-\bar{\zeta}~{\rm sector},~G_{\mu\bar{\nu}}&=&l^2 {\delta_\mu}^\nu C'/2.
\end{eqnarray}
In the most general situation, expressions $C,C'$ represent polynomials of permitted 
{\em gauge-invariant} curvature terms:
\begin{equation}
C \equiv 1 + \sum_{r=1}^N c_r Z^r,\qquad C' \equiv 1 + \sum_{r=1}^N c'_r Z^r;
       \quad Z\equiv \bar{\zeta}\zeta.
\end{equation}
The inverse metric components are readily found:
\begin{eqnarray}
x-x~{\rm sector,}~G^{mn}&=&g^{mn}C^{-1},\\
x-\zeta~{\rm sector,}~G^{m\nu}&=& i(A^m\zeta)^\nu C^{-1},\\
\zeta-\bar{\zeta}~{\rm sector,}~G^{\mu\bar{\nu}}&=& [2{\delta_\nu}^\mu/l^2
-(\bar{\zeta}A^m)^{\bar{\nu}}(A_m\zeta)^\mu]\, C'^{-1}.
\end{eqnarray}
A general rotation of the property coordinates, $\zeta \rightarrow \exp[i\Theta(x)]\zeta$\,
then just corresponds to a gauge transformation of the force fields, so we anticipate
that the Ricci superscalar $\cal R$ must turn out to be gauge invariant. Indeed it is,
as verified multiple times through Mathematica evaluations.

To further the calculation of the dependence of the Lagrangian on the property curvature 
coefficients $c_r$ and $c'_r$, we will require the Berezinian\cite{B,W} of the metric.
The latter is actually gauge invariant so we can set $A \rightarrow 0$ to evaluate it. 
A simple calculation produces
\begin{equation}
\sqrt{G..} =\sqrt{g..} C^2 (l^2 C'/2)^{-N}.
\end{equation}
We will also need the Grassmann integrals,
\begin{eqnarray}
\int(d^N\!\zeta\,d^N\!\bar{\zeta})\,\, Z^N &=& (-1)^{\langle N\rangle} N!,\\
\int(d^N\!\zeta\,d^N\!\bar{\zeta})\,\,Z^{N-1} (\bar{\zeta}H\zeta)
   &=&(-1)^{\langle N\rangle} (N-1)!\,{\rm Tr}\,H,
\end{eqnarray}
where $\langle M\rangle$ signifies int[$M$/2] and $H$ stands for 
a general U($N$) matrix  in property space.

Before continuing, we will find it useful to use alternative parametrizations to (6), since
they help to simplify the ensuing analysis; namely write
\begin{equation}
C=\exp[-\sum_{r=1}^N a_r\,Z^r],\qquad C'=\exp[-\sum_{r=1}^N a'_r\, Z^r]; \quad Z\equiv\bar{\zeta}\zeta.
\end{equation}
Thus derivatives are easily found,
\begin{equation}
\frac{\partial C}{\partial \zeta} = \sum_{r=1}^N \bar{\zeta}ra_r Z^{r-1}\,C\equiv\bar{\zeta}DC,\quad
\frac{\partial C}{\partial \bar{\zeta}} = -\sum_{r=1}^N ra_r Z^{r-1}\,\zeta\,C,\equiv -D\zeta C,
{\rm ~~etc.}
\end{equation}
Of course there is a simple translation table between parametrizations (6) and (13):
\begin{eqnarray}
a_1&=&-c_1,\quad a_2=c_1^2/2 - c_2, \quad a_3 = -c_1^3/3 +c_1c_2 - c_3,\\
a_4&=&c_1^4/4-c_1^2c_2+c_2/2+c_1c_3-c_4,\\
a_5&=&-c_1^5/5+c_1^3c_2-c_1c_2^2-c_1^2c_3+c_2c_3-c_5,\quad{\rm etc.}
\end{eqnarray}
In this $a$-parametrization, $\sqrt{G..} =\sqrt{g..}\,(2/l^2)^N\exp[\sum_{r=1}^N (Na'_r-2a_r)Z^r]$.

\section{Determination of the various parts of the Lagrangian}
We are after the integral of the superscalar curvature, 
$\int(d^N\!\zeta\,d^N\!\bar{\zeta})\,\sqrt{G..}\,{\cal R}$, which will produce three sorts of terms:
the purely gravitational bit $R^{[g]}$, the gauge contribution proportional to Tr $F.F$, where
$F$ is the generalised curl of the gauge field, and finally the constant, cosmological part.
The tactic is to identify relevant bits of each by picking out appropriate pieces of
$G^{KL}G^{MN}G^{RS}(\partial_.G_{..}\partial_.G_{..})_{KLMNRS}$.

\subsection{The gravitational term}
The gravitational curvature arises from the structure $g^{kl}g^{mn}g^{rs}
(\partial.g..\partial.g..)_{klmnrs}$, which itself comes from $G^{kl}G^{mn}G^{rs}
(\partial.G..\partial.G..)_{klmnrs}$ and therefore carries the factor $C^{-1}$ as can be
ascertained from eqs. (3) and (7). Including the Berezinian, we see that
\begin{equation}
\sqrt{G..}{\cal R} \supset \sqrt{g..}\,R^{[g]}\,(2/l^2)^N\,\exp\left[\sum_1^N (Na'_r-a_r)Z^r\right].
\end{equation}
Upon $\zeta$ integration we deduce that
\begin{equation}
\int(d^N\!\zeta d^N\!\bar{\zeta})\,\sqrt{G..}{\cal R} \supset \sqrt{g..}\,R^{[g]}\,
 (-1)^{\langle N\rangle} \left(\frac{2}{l^2}\frac{d}{dZ}\right)^N\exp\left[\sum_1^N (Na'_r-a_r)Z^r\right]
  \bigg |_{Z=0}
\end{equation}
that we can later convert into $c$-form, if desired.

\subsection{The gauge field term}
The  curl $F_{mn}=A_{n,m}-A_{m,n}+i[A_n,A_m]$, which we are sure arises in the Lagrangian,
can be be picked out by focussing on the first derivative of the gauge field and ignoring 
the other parts as these will  come automatically. 
Now the gauge field occurs in the $x-\zeta$ component $G_{m\nu}$ of the
metric and is attached to a factor of $\zeta$ as well as $C'$. Therefore we need only
examine terms of the type  $G^{km}G^{ln}G^{\sigma\bar{\rho}}(\partial_kG_{l\bar{\rho}}
\partial_mG_{n\sigma})$ and these engender a factor $(l^2C'/2)C^{-2}(\bar{\zeta}F.F\zeta)$
upon contraction over indices. It follows that, apart from a proportionality factor,
\begin{equation}
\sqrt{G..}{\cal R}\supset\sqrt{g..}\,(2/l^2)^{N-1}(\bar{\zeta}F.F\zeta)
\exp\left[\sum_{r=1}^N (N-1)a'_rZ^r\right].
\end{equation}
Integration over property (see eq. (12)) yields
\begin{equation}
\int(d^N\!\zeta d^N\!\bar{\zeta})\sqrt{G..}{\cal R} \propto \sqrt{g..}{\rm Tr}(F^{mn}F_{mn})
      (-1)^{\langle N\rangle}\!\left(\frac{2}{l^2}\!\frac{d}{dZ}\right)^{\!\!N-1}
      \!\!\!\!\!\!\!\exp\left[\sum_{r=1}^N (N\!\!-\!\!1)a'_rZ^r\right]\!\bigg |_{Z=0}.
\end{equation}
One readily checks via the case $N=1$ that the proportionality factor needed is just -1/2.

\subsection{The cosmological term}
This piece, which like $\sqrt{G}$ does not depend on $A$, is a bit more complicated because 
it can arise from three types of contribution:
\[
(G^{kl}G^{mn} ~{\rm or}~G^{km}G^{ln})\,G_{kl,\rho}G_{mn,\bar{\sigma}}G^{\bar{\sigma}\rho},
\] 
\[
(G^{kl}G_{kl,\bar{\mu}}G_{\rho\bar{\sigma},\nu})\,(G^{\nu\bar{\mu}}G^{\bar{\sigma}\rho}~{\rm or}~
   G^{\rho\bar{\mu}}G^{\bar{\sigma}\nu})
   \]
\[
(G^{\kappa\bar{\lambda}}G_{\bar{\lambda}\kappa,\bar{\mu}}G_{\rho\bar{\sigma},\nu})
  (G^{\nu\bar{\mu}}G^{\bar{\sigma}\rho}~{\rm or}~   G^{\rho\bar{\mu}}G^{\bar{\sigma}\nu}).
\]
Each of these has to be taken with with multiplicative factors $\alpha_N,\beta_N,\gamma_N$
respectively and may depend on $N$ through the contraction 
$G^{\bar{\rho}\nu}G_{\nu\bar{\rho}}\propto N$. Thus we can ascertain 
via comparison with the $N=1,2$ and 3 cases that $\alpha_N$=6 
is $N$-independent, $\beta_N$ is linear in $N$, namely $\beta_N=-4(2N+1)$;  and that $\gamma_N=(2N+1)(N+1)$ is $N$-quadratic. 
(Alternatively $\alpha,\beta,\gamma$ can be painstakingly determined from first principles.)
Now
\[
G^{kl}G^{mn}G_{kl,\rho}G_{mn,\bar{\sigma}}G^{\bar{\sigma}\rho}~{\rm entrains}~
   (2/l^2)\,C'^{-1}C^{-2}\left(Z\frac{dC}{dZ}\frac{dC}{dZ}\right),
\]
\[
G^{kl}G_{kl,\bar{\mu}}G_{\rho\bar{\sigma},\nu}G^{\nu\bar{\mu}}G^{\bar{\sigma}\rho}
~{\rm entrains}~   (2/l^2)\,C'^{-2}C^{-1}\left(Z\frac{dC}{dZ}\frac{dC'}{dZ}\right),
\]
\[
G^{\kappa\bar{\lambda}}G_{\bar{\lambda}\kappa,\bar{\mu}}G_{\rho\bar{\sigma},\nu}
~{\rm entrains}~   (2/l^2)\,C'^{-2}C^{-1}\left(Z\frac{dC'}{dZ}\frac{dC'}{dZ}\right).
\]

Altogether we can conclude that the cosmological term arises from
\begin{equation}
\sqrt{G..}{\cal R}\supset\sqrt{g..}\,(2/l^2)^{N\!+\!1}Z.\exp[\sum_{r=1}^N((N+1)a'_r\!-\!2a_r)Z^r]\,
       [\alpha_ND^2\!+\!\beta_N DD'\!+\!\gamma_N D'^2],
\end{equation}
with
\[
\alpha_N=6,\quad \beta_N=-4(2N+1),\quad \gamma_N=(2N+1)(N+1).
\]
All that is left is to integrate (12) over property.

\subsection{The full result for any $N$}
Upon $\zeta$ integration we end up with the totality,
\begin{eqnarray}
(-1)^{\langle N\rangle}&&\!\!\!\!\left(\frac{l^2}{2}\right)^{\!\!\!N}\!\!\!\int\!(d^N\!\zeta d^N\!\bar{\zeta})
\,\sqrt{G..}{\cal R}
= \sqrt{g..}(\frac{d}{dZ})^N\,\left(R^{[g]}\,{\rm e}^{\sum(Na'_r-a_r)Z^r}\right.\nonumber\\
&& -\frac{l^2}{4N} {\rm Tr}F.F\,Z \,{\rm e}^{\sum(N-1)a'_rZ^r}\nonumber \\
&& +\left. \!\!\frac{2Z}{l^2}{\rm e}^{\sum((N\!+\!1)a'_r\!-\!2a_r)Z^r}
\!\!\{6D^2\!-\!4(2N\!+\!1)DD'\!+\!(2N\!+\!1)(N\!+\!1)D^2\}\!\!\right)\!\!\bigg|_{Z=0}.
\end{eqnarray}
Converting from $a_r$ to $c_r$ via (15) to (17), the reader can verify that the 
results for $N=1,2,3$ stated in previous papers emerge correctly.  These are tabled below.

\begin{table}[h]
\tbl{Coefficients of terms multiplying gravity, gauge and cosmological pieces (up to $N=3, C=C'$)}
{\begin{tabular}{@{}cccc@{}} \toprule
$N$ & $R^{[g]}$ &Tr $F.F$ &cosmological constant \\
\colrule
1 & $(2/l^2)(c_1-c'_1)$ & -1/2 & $(24/l^4)(c_1-c'_1)^2$ \\
2 & $(8/l^4)(2c_1c'_1\!-\!3{c_1}'^2\!-\!c_2\!+\!2c_2')$ & $-c'_1/l^2$ & 
     $-(16/l^6)(24c_1c_2\!-\!38{c_1}^2c'_1\!-\!40c_2c_1'\!+$ \\
 \hphantom{0}&\hphantom{0}&\hphantom{0}&$110c_1{c_1}'^2\!-\!75{c_1}'^3\!-\!40c_1c_2'\!+\!60c_1'c_2')$\\   
3 &$(96/l^6)(2{c_1}^3-3c_1c_2+c_3)$& $(4/l^4)(3{c_1}^2\!-\!2c_2)$ & 
$-(1152/l^8)(5{c_1}^4\!-\!10{c_1}^2\!+\!2{c_2}^2\!+\!3c_1c_3)$\\
\botrule
\end{tabular}\label{ta1} }
\end{table}
In this way, the coefficients can be fully determined analytically for any $N$ {\em without resorting 
to algebraic computer packages}.

\section{Application to $N=4$}
We may apply the  above technique to the case where charge and colour 
(electricity and chromicity properties) are taken together. Since QED and QCD are
parity invariant we need not concern ourselves with different properties for handedness,
as one would need to for electroweak theory. Let indices on $\zeta$ of $i=1,2,3$ refer to colour 
(red, green, blue, carrying charge 1/3 as per $\bar D$ quarks)  and 4 refer to electronic 
charge (-1). This means combining the coupling of the gluon fields $B_m$ with the electromagnetic
field $A_m$ in the property-spacetime sector:
\begin{equation}
G_{m4} =il^2\zeta^{\bar 4}\,eA_m/2;\quad 
G_{mi} =il^2[-\zeta^{\bar i}\,eA_m/3 +\zeta^{\bar j}\,{fB_m}^{j \bar i}]/2.
\end{equation}
The burning issue is whether we are forced to assume that the couplings $e$ and $f$ must
merge at some high energy scale in order to ensure gravitational universality (as we needed
to when discussing chirality) or whether there is sufficient freedom allowing them to differ
from each other. In fact we shall presently see that the latter holds.

To that end we will {\em simplify} the argument by adopting an overall set of curvatures $C=C'$
arising in (3) to (5). However, because we are dealing with a direct product U(1)$\times$SU(3) 
gauge group, we have at our disposal two independent property invariants: 
$\zeta^{\bar 4}\zeta^4$ and $\zeta^{\bar i}\zeta^i$. Of particular interest is the possibility of
\[
C=1+\ldots+c_e(\zeta^{\bar 4}\zeta^4)(\zeta^{\bar i}\zeta^i)^2 +c_f(\zeta^{\bar i}\zeta^i)^3+\ldots,
\]
involving two curvature constants $c_e$ and $c_f$. As we are interested in the gauge field
contributions to the Lagrangians, we must focus on terms having the structure
$G^{\mu{\bar \nu}}G^{kl}G^{mn}(\partial_mG_{k\bar{\nu}})(\partial_nG_{l\nu})$, which entrain a
overall factor $C^{-1}$ multiplying the flat field case. In particular we find that
\begin{eqnarray}
G^{4{\bar 4}}G^{kl}G^{mn}(\partial_mG_{k\bar{4}})(\partial_nG_{l4})&\!\rightarrow\!&
   g^{km}g^{ln}e^2\zeta^{\bar 4}F_{kl}F_{mn}\zeta^4 \nonumber \\
G^{i{\bar j}}G^{kl}G^{mn}(\partial_mG_{k\bar{j}})(\partial_nG_{li})&\!\rightarrow\!&   
    g^{km}g^{ln}[e^2\zeta^{\bar 4}F_{kl}F_{mn}\zeta^4/3 \!+\!
                           f^2\zeta^{\bar i}(\!E_{kl}E_{mn}\!)^{i{\bar j}}\zeta^j],
\end{eqnarray}
where $F_{mn}\equiv A_{n,m}-A_{m,n}$ and $E_{mn}\equiv B_{n,m}-B_{m,n}+if[B_n,B_m]$
are the standard ``curls'' of the electromagnetic and colour fields respectively.

Now remembering that for four properties $\sqrt{G..}=(2/l^2)^4\sqrt{g..}C^{-2}$ we deduce that
the sum of the gauge field contributions will be held in the expression
\begin{eqnarray}
{\cal R}\sqrt{G..}&\supset&
[1-3c_e(\zeta^{\bar 4}\zeta^4)(\zeta^{\bar i}\zeta^i)^2 -3c_f(\zeta^{\bar i}\zeta^i)^3+\ldots].\nonumber\\
 & &\,\,g^{km}g^{ln}[4e^2\zeta^{\bar 4}F_{kl}F_{mn}\zeta^4/3 \!+\!
                           f^2\zeta^{\bar i}(\!E_{kl}E_{mn}\!)^{i{\bar j}}\zeta^j].
\end{eqnarray}
It only remains to integrate over the four properties to discover the gauge field Lagrangian,
(including appropriate factors of $l^2$)
\begin{equation}
\int(d^4\!\zeta d^4\!\bar{\zeta})\sqrt{G..}{\cal R}\supset
 -(12/l^2)[4c_fe^2F.F + c_ef^2{\rm Tr}(E.E)].
\end{equation}
Clearly all one needs to do is to set $c_ef^2=4c_fe^2$ and we maintain a uniform
gravitational constant in the ensuing work, {\em without forcing equality of the 
colour and electromagnetic couplings}. Relaxing the assumptions $C=C'$ and the form
of $C$ makes it even easier to ensure uniformity of $G_N$.

\section{Conclusions}
We have described a `first-principles' way of determining the Ricci coefficients for 
spacetime curvature, property curvature (embodied in the cosmological constant) 
and gauge field Lagrangians, which arise
from the super-Ricci scalar. The method represents a major advance as it unshackles us from
relying on a computer algebra package, which struggles timewise as the 
number of properties rises. The final results just depend on the property curvature coefficients
which enter the supermetric while maintaining gauge covariance and they have been listed in
Table 1. The procedure puts us in a strong position for handing electroweak theory and the full standard model unification with gravity.

\end{document}